\documentclass[review,number,sort&compress]{elsarticle}
\journal{Nuclear Instruments and Methods A}
\begin{document}

\begin{frontmatter}
%\linenumbers

\title{ Pulse shape analysis of neutron signals in Si-based detectors}

\author[aut1,aut3] {G. Mauri,}
\author[aut1]{M. Mariotti,}
\author[aut2]{F. Casinini,}
\author[aut1,aut2]{F. Sacchetti,}
\author[aut1,aut2,aut4]{C. Petrillo}

\address[aut1]{Dipartimento di Fisica e Geologia, Universit\'a di Perugia, Via A. Pascoli, I-06123 Perugia, Italy}
\address[aut2]{Istituto per l'Officina Materiali of CNR, Via A. Pascoli, I-06123 Perugia, Italy}
\address[aut3]{European Spallation Source ERIC (ESS), P.O. Box 176, SE-22100 Lund, Sweden.}
\address[aut4]{INFN, Sezione di Perugia, Via A. Pascoli, I-06123 Perugia, Italy}

\begin{abstract}

A series of test experiments on three different Si-based neutron detectors, namely a 1-d 128 channel, 0.5 mm space resolution Si microstrip sensor coupled to natural Gd converter, a medium size $\sim$ 1 cm$^2$ PIN diode coupled to $^n$Gd$_2$O$_3$ or $^{157}$Gd$_2$O$_3$ converters, and a SiPM photomultiplier coupled to neutron scintillators, are presented to show the performances of this class of devices for thermal neutron detection. A pulse shape analysis method, designed to improve the performance of these devices by an optimized discrimination of the neutron signals from noise and background radiation, is proposed, described and tested. This study is aimed to real time applications and single event storage of the neutron information in time of flight instrumentation.

\end{abstract}

\end{frontmatter}
%\linenumbers
\section{Introduction}

The enormous growth of the experimental possibilities~\cite{iop} at the neutron sources, despite the limited increase of the source brightness since the 1960s, has been enabled by the development of instrumentation like large position-sensitive detectors, new focusing optics and innovative exploitation of neutron polarization. The operation of new generation, high-intensity neutron sources like the Spallation Neutron Source~\cite{sns} (Oak Ridge Laboratory, TN, USA), the Japan Spallation Source~\cite{jsns} at J-PARC (Tokai-mura, Japan), and the European Spallation Source (ESS, Lund, Sweden), presently under construction, where an intensity increase of at least an order of magnitude is expected~\cite{ess1,ess2,ess3}, represents a challenge because of the increased performance demands on neutron detectors. For example, the typical process of charge collection in the widely used $^3$He gas detectors is in the $\mu$s range, which introduces appreciable systematic uncertainties on the detector dead time and the signal pile-up or errors at detection rates higher than 100 kHz. This is not of primary concern for most applications at the present installations, whereas it can become rather important in the near future when the new high rate instrumentation of the ESS will start operation. There, new technologies will be relevant to achieve better performances in neutron detection time, below 1 $\mu$s and possibly tens of ns for some specific applications.

Overall reviews of the progress made in neutron detectors are available~\cite{caru,epj} while the search of new solutions is continuous, driven by the needs of new instruments and the anticipated shortage of $^3$He. Alternative approaches, exploiting $^{10}$B-based solutions, have been recently proposed~\cite{multi} as large area thermal neutron detectors for high counting rates applications. This solution will find wide application at the ESS, although not typically aimed at high space resolution usage. Lately, the development of new devices based on solid boron nitride crystals has been reported~\cite{bn} and their application proved. It is reasonable to expect novel concepts to translate into new detector applications in the near future, coupled to improved data collection techniques as prompted by an extended use of time of flight techniques.

The requirements for large area, and simultaneously flexible shape, detectors with sub millimetric resolution could be satisfied by resorting to the technology based on solid-state Si devices. Prototypes of solid state neutron detectors were proposed in the past decades~\cite{n1,n2,str} and the possibility of reaching sub millimeter resolution\cite{n2} with adequate detection efficiency~\cite{n1,n2} was proved~\cite{caru,epj}. Implementation of silicon detectors for neutron applications has the advantage of exploiting Integrated Circuit (IC) technology, which makes possible to meet the demand for quite high-density readout electronics. Silicon detectors can be considered as a robust and mature alternative, already available on the market, and characterized by a detection efficiency comparable with that of the current solutions~\cite{multi} at least over the thermal neutron region. The high space resolution at high counting rate, typical of the Si p-n junction diode, together with the operability under vacuum and in high magnetic fields, are most appealing features of a Silicon detector for operation under the intense neutron fluxes expected at the future installations. Very interesting indeed is the maximum instantaneous data rate enabled by the coupled use of Si detectors and VLSI fast electronics. Operating the detector as a counter under an intense and pulsed neutron beam actually requires a dead time notably less than 1 $\mu$s and the ability of managing the time information contained in the incoming neutron pulse. Readout of a large number of channels, at high speed rate and in parallel mode, is therefore a fundamental requisite for the construction of a large area detector.

The currently available VLSI front-end electronics can reach peaking times as short as 50 ns with still high gain ($\sim$30 mV/fC) and low noise (less than 500 e$^-$). It is so possible to get an instantaneous readout speed up to several MHz per channel. Nonetheless, flexibility and reliability of the data collection need to be implemented at the highest possible level. Over the last years, large effort was spent in designing and testing new neutron detectors and in developing better approaches to discriminate neutron signals against the background radiation, as well as to analyze the detector and electronic noise. In particular, several pulse shape and real time analysis approaches were proposed~\cite{ps1,ps2,ps3,psb4,psb5,psb6}, including the use of Field Programmable Gate Array (FPGA) to perform real time operations on the detected signals. In a recent paper~\cite{n3} we reported on the use of an FPGA-based system for single-event time of flight (ToF) applications as a practical tool to eliminate the time uncertainty on the collection time at the detector.

Here we present several series of neutron scattering data on standards, which we obtained by different Si devices and prototype detectors, and the results of signal processing by pulse shape analysis designed to extract the neutron signal from high noise, in real time mode. We first discuss the data collection using Si microstrip devices, then the neutron tests, to finally introduce the pulse shape analysis, efficient and simple enough for real time applications using FPGAs.

\section{Silicon microstrip position sensitive neutron detector}

Solid state neutron detectors based on silicon microstrips have been tested to show that sub-millimeter spatial resolution together with a rather high counting rate can be easily reached using the now mature technology of integrated silicon devices. The device here employed exploits the same design as presented in Ref.~\cite{n3}. It is a 1-d 0.55 mm resolution system built on a Si microstrip with 110 $\mu$m pitch and a total surface of about 70 x 40 mm$^2$. The backplane of the microstrip is directly coupled to a natural Gd plate, 0.25 mm thick, which operates as the full neutron absorber. The solid state silicon sensor, the microstrip, detects neutrons captured inside the Gd plate through the charge released inside silicon by the secondary electrons. Due to the relatively low electron energy, the charge collection in silicon is rather small, ranging from 1 fC to 10 fC, therefore a low noise front-end electronics must be employed. The charge collected by the silicon microstrip sensor is integrated and shaped using an IDEAS, Norway~\cite{ideas} VA32 IC which provides 32 channels with a nominal peaking time of 75 ns. The shaped signals are then discriminated and converted into a digitally coded output using an IDEAS TAN32 IC which allows for a window discrimination of the analog pulses. The digital signals generated by the TAN IC provide both the channel address and the time stamp. These signals are used to store the full neutron information with a time resolution better than 100 ns.

To define the performances of this detector a series of neutron tests were performed on the ALF beam line at ISIS pulsed neutron source (Chilton, Didcot, UK) by directly coupling the system to the Data Acquisition Electronics of the facility. The ALF beam line is located on target 1 downstream another instrument (PRISMA) and the intensity available for the neutron tests was limited by the presence of the sample environment of the main experiment. Additional measurements were carried out at the IN3 beam line of the ILL steady source (Grenoble, France) with the purpose of testing the single event acquisition electronics that was coupled to the neutron detector.

\begin{figure}
%\vspace{20mm}
\includegraphics[width=0.9\textwidth,keepaspectratio]{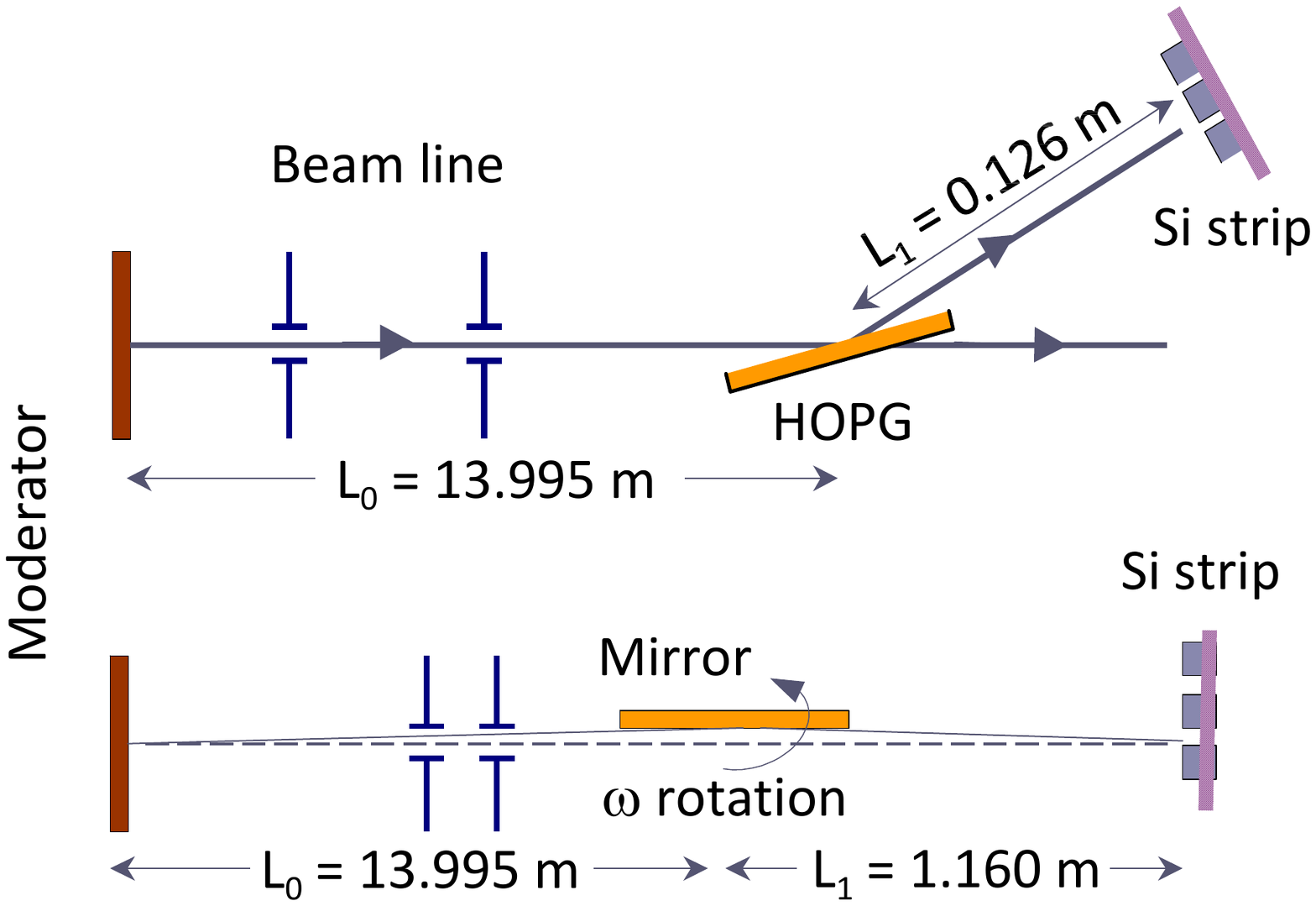}
\caption{Schematics of the neutron measurements setup on the ALF beam line at ISIS. The upper panel refers to the HOPG crystal or the Ni polycrystal configuration while the lower panel refers to the reflectivity setup for the test on the mirror. Note the short secondary path used in both cases.}
\label{fig1}
\end{figure}

\begin{figure}
%\vspace{20mm}
\includegraphics[width=0.45\textwidth,keepaspectratio]{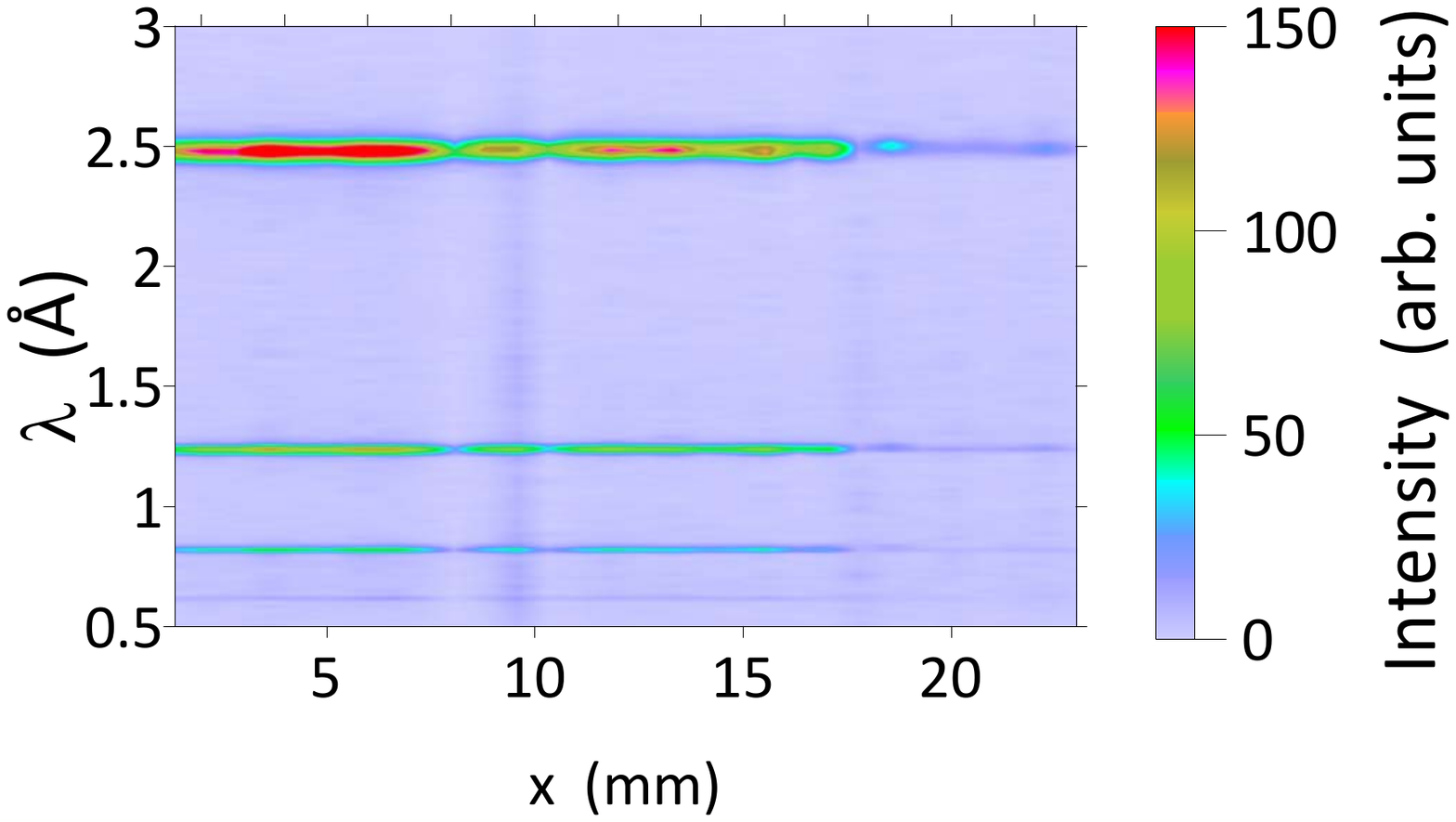}
\includegraphics[width=0.45\textwidth,keepaspectratio]{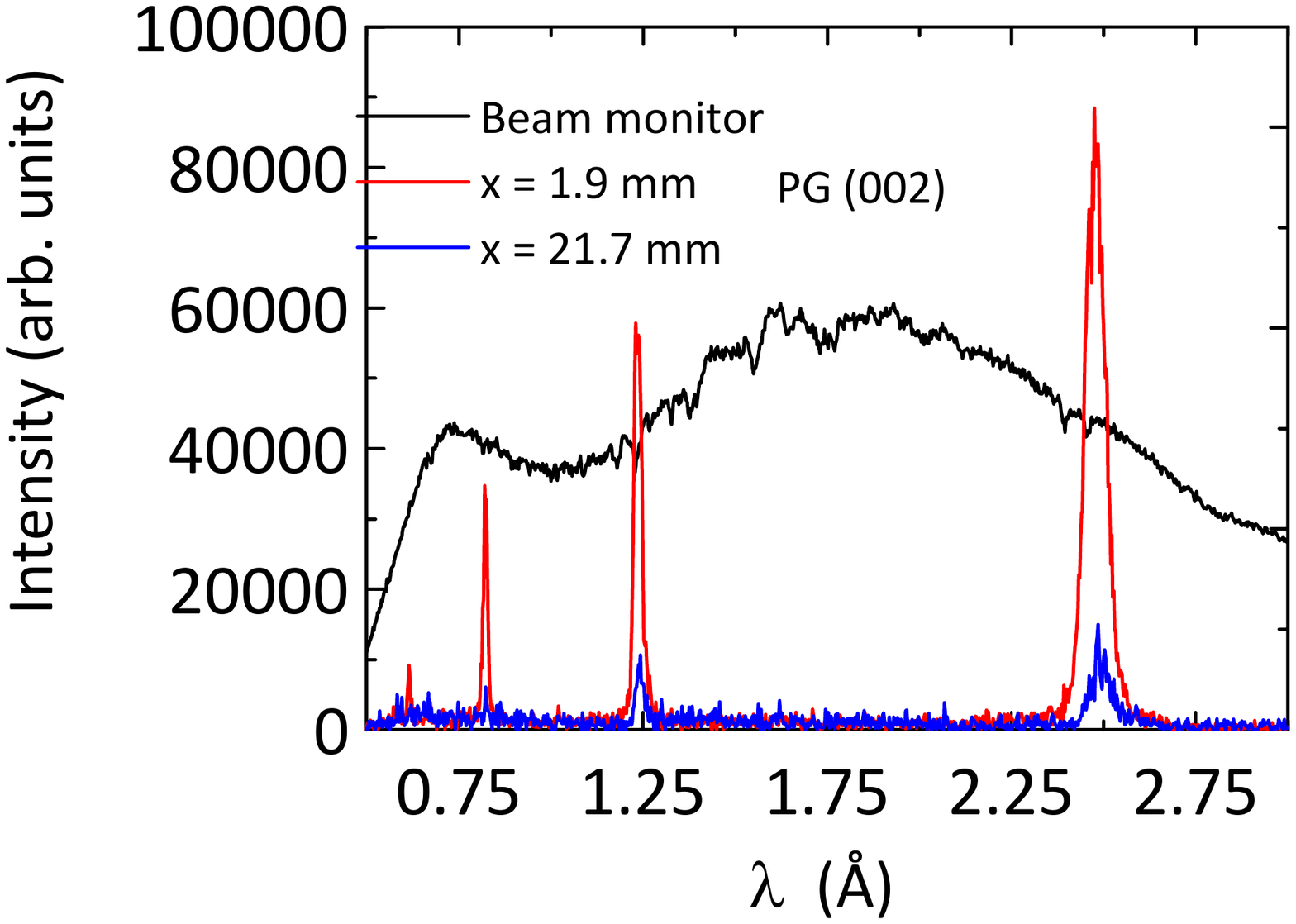}
\caption{HOPG crystal sample. Panel a), diffracted intensity as a function of position and neutron wavelength. From top to bottom the four reflections (002), (004), (006) and (008) are visible at $\lambda$ equal to 2.48 \AA, 1.24 \AA, 0.83 \AA, and 0.62 \AA. Panel b), intensity versus wavelength collected on the fourth strip located at about 1.9 mm from the sensor edge (red line) and on the 40-th strip located at about 21.7 mm from the sensor edge (blue line), practically out of the diffracted beam. The intensity measured by the beam monitor is also shown for comparison (black line).}
\label{fig2}
\end{figure}

The first series of measurements was carried out on a highly oriented pyrolitic graphite (HOPG) crystalline sample, 70 mm long and 40 mm high, 0.5$^{\circ}$ mosaic spread, oriented at an angle 21.7$^{\circ}$ off the incoming beam. A B$_4$C window, 1 cm thick and 6 mm wide, was mounted to define the divergence of the incoming beam, which is almost parallel because of its $\sim 0.3^{\circ}$ intrinsic collimation. The HOPG crystal was mounted at the sample position of the beam line and the detector was aligned with its geometrical center 126 mm from the sample position and set at $\sim 40^{\circ}$ from the beam direction. The crystal was only partially illuminated by the incoming beam. The test configuration is sketched in Fig.~\ref{fig1} where it can be observed that the detector covered a range of about 15 mm. Apart from Cd sheets on the sides of the detector, no further shielding was employed as the only expected background source was from the air scattering. In Fig.~\ref{fig2} the intensity collected by the strips 4 and 40 (1.9 mm and 21.7 mm from the crystal edge) is shown in comparison to the monitor counts. A clean diffraction pattern of HOPG is obtained up to the (008) reflection, on the fourth strip, after counts integration up to $\sim 100 \mu$A h on the proton target. Strip 40 at 21.7 mm from the edge of the sensor was almost outside the incoming beam and the intensity dropped abruptly from strip 41 onwards.

\begin{figure}
%\vspace{20mm}
\includegraphics[width=0.9\textwidth,keepaspectratio]{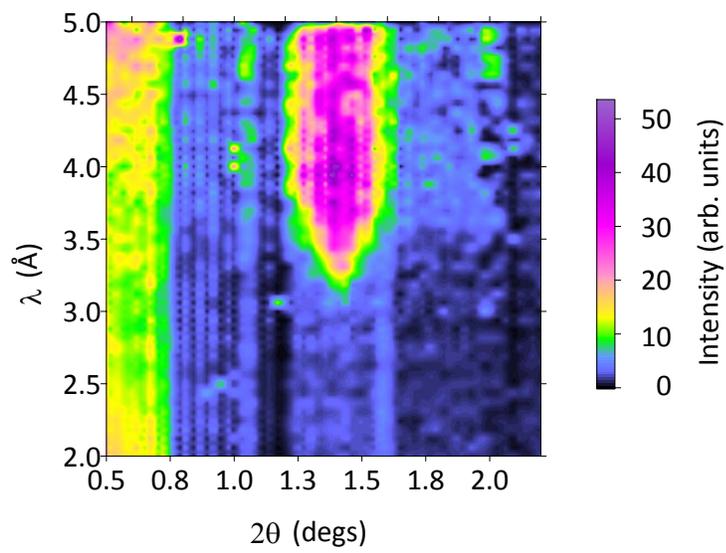}
\caption{Intensity map diffused by the mirror located at the sample position as a function of diffraction angle at the detector and wavelength. The critical wavelength at about 3.4 \AA\ is clearly visible.}
\label{fig3}
\end{figure}

\begin{figure}
%\vspace{20mm}
\includegraphics[width=0.9\textwidth,keepaspectratio]{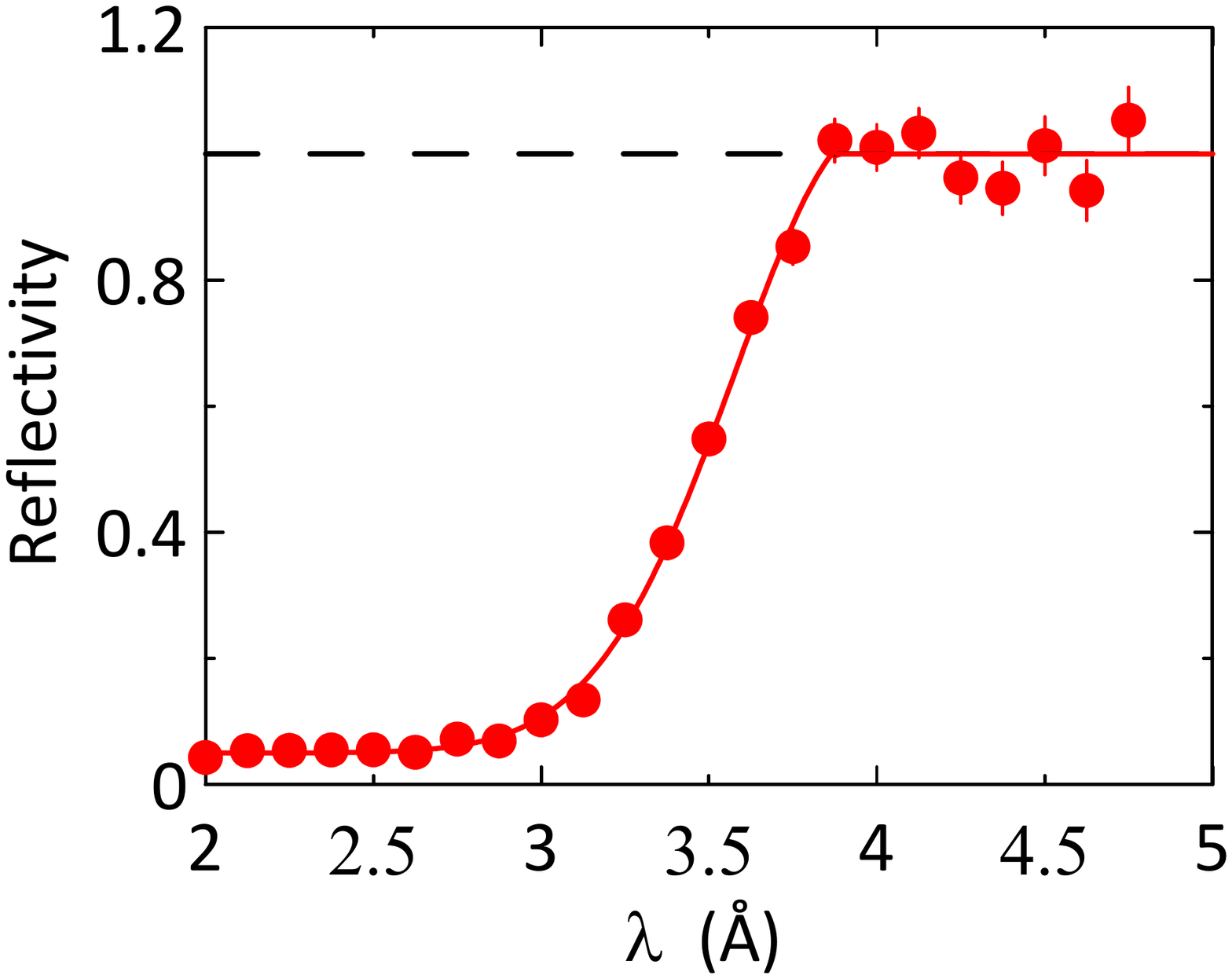}
\caption{Reflectivity curve of the mirror collected integrating the diffracted intensity on four strips corresponding to about 0.1 degree interval. The raw data are shown.}
\label{fig4}
\end{figure}

The second test was aimed at measuring the performances at small angle scattering conditions, and a 3 mm wide beam was prepared to impinge on an $m = 2$ supermirror to collect the totally reflected neutrons.  The detector was installed by the instrument beam stop at about 1160 mm from the mirror that was mounted on a rotating stage at the sample position. Fig.~\ref{fig3} shows the diffused intensity after collection up to $\sim 250 \mu$A h on the proton target. The incoming neutron beam had a divergence of about 0.3$^{\circ}$, which must be compared with the nominal angular width related to the beam size and distribution. Under the simple assumption of a triangular distribution, a beam image at the detector of about $\pm 5$ mm width from to the center is expected. By this estimate, one can guess that the beam extends up to $0.4^{\circ}$ degrees from the nominal zero position. To avoid overloading the acquisition system by the direct beam intensity, a 10 mm wide B$_4$C shield was inserted to stop the straight beam. The major source of background is the air scattering taking place along the long flight path, and the tail of such a scattering is clearly visible in Fig.~\ref{fig3}. At $\sim 1.4^{\circ}$, the beam reflected from the mirror starts appearing at long enough wavelengths. Also, as expected, the reflection occurs at longer wavelength when the reflection angle is different from the ideal one. To directly examine the reflectivity curve, we collected the data at $1.40^{\circ} \pm 0.03^{\circ}$ as a function of the wavelength. The result is shown in Fig.~\ref{fig4} where the correct behavior of the mirror is apparent. Although the background is, unfortunately, too high to properly get the reflectivity curve, the overall trend of the data indicates that, by means of a proper shielding and evacuating the flight path, this kind of detector can be adequately applied for reflectivity measurements.

\begin{figure}
%\vspace{20mm}
\includegraphics[width=0.9\textwidth,keepaspectratio]{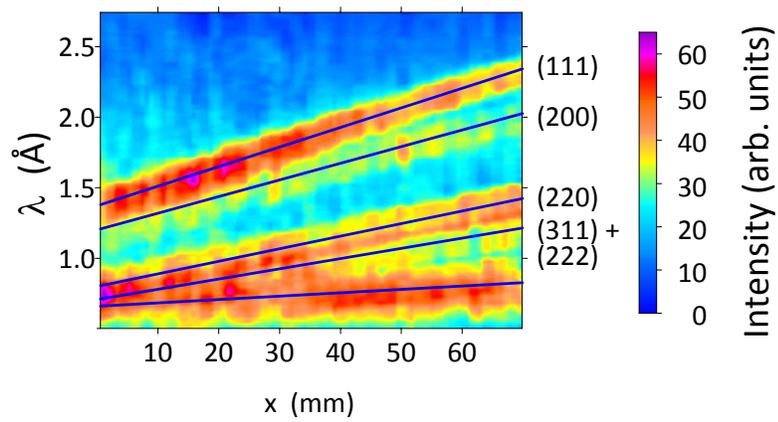}
\caption{Intensity map of the 3 mm diameter Ni sample. The two dimensional plot does not allow for the resolution of the very short wavelength reflections below 1 \AA.} 
\label{fig5}
\end{figure}

The third series of measurements consisted in collecting the powder diffraction pattern from a Ni rod, 3 mm diameter. The experiment was carried out in the same configuration as the HOPG crystal diffraction and the data were collected corresponding to a total charge on the target of about 2500 $\mu$A h. The overall diffraction pattern is plotted in Fig.~\ref{fig5} where the Ni Bragg reflections are shown as a function of the scattering angle.

All the above tests were performed by successfully transferring a time stamp signal and the digital address of each strip directly to the acquisition electronics available at ISIS, so enabling access and storage of the data within the main data acquisition (DA) system of the facility. The ISIS DA electronics distributes the neutron counts into time bins according to their arrival time, from which counts can be transformed into more practical distributions, like neutron wavelength and neutron energy. With small enough time bins, this system provides a working and sensible way to collect the data. The only limitation is, however, the presence of a relatively large number of empty bins when the intensity is low. To obtain a (${\bf Q}, \omega$) space map requires collection of all the data and conversion of time bins into energy bins on different detectors by the proper transformation that weights each neutron according to its original time and detector position, {\sl i.e.} strip. This procedure introduces correlations between different bins on the energy axis, for instance, as the same time bin can contribute partly to distinct energy bins.

Given the ISIS DA architecture, we designed and used an acquisition system based on a single event capture~\cite{n3}. Each neutron collected on a given detecting strip is stored as a four bytes unsigned integer. The bits are distributed to provide a high-resolution time stamp with respect to a user-provided trigger, e.g. the nominal arrival time of the protons on the target of the short pulse source, and a number to identify the strip where the neutron has been absorbed. The acquisition system, based on FPGA and described in Ref.~\cite{n3},  was also used in combination with rotating systems for ToF instrumentation. In the case of the rotating crystal monochromator~\cite{rotat} there is quite some advantage in driving both the rotating device and the detector acquisition by means of a single FPGA, as the phasing of the ToF acquisition is automatically obtained since all the FPGA activity is based on a single clock. The stability of this phasing system is almost perfect and no phasing adjustment is necessary. A test on a rotating crystal monochromator~\cite{rotat}) was carried out on the IN3 beam line of the ILL.

\section{Real time pulse shape analysis of the neutron signals}

Si sensors coupled to Gd converters can be used as solid state neutron detectors in typical condensed matter experiments with acquisition systems effective also for other PSD detectors, as it has been the case of the experiments performed with the ISIS DA. However, a more compliant acquisition electronics would be desirable for the specific needs of a neutron counter for low energy scattering applications. In particular, some flexibility in optimising the real time signal analysis can be crucially relevant. Indeed, the ability to operate also in high noise environment, where the low signal from Si detectors is a limitation, would be a major step forward for this technology. To this purpose we developed a new approach for a real time analysis of the detector pulses.

In a standard acquisition system the front end of the detector is a charge preamplifier, with appropriate filters, followed by a shaper amplifier, a simple threshold discriminator to produce a short pulse used to determine the arrival time of a neutron. In presence of noise, the discriminator threshold is set higher and part of the neutron counts can go lost while, on the contrary, some noise pulses can be counted as neutrons. This kind of problems are more severe for solid state Si detectors where the charge signal is rather small. For a more accurate identification of neutrons, it is also preferable to employ a window discriminator.

The general principle of the acquisition system turns out to be completely different when using an FPGA connected to the very front-end of the detector. The basic design we implemented consists of a front-end wide band and low noise amplifier plus an Analog to Digital Converter (ADC) with a frequency appropriate for a proper sampling of the signal produced by the detector in use. The digital output of the $i$-th ADC is sent directly to the FPGA to $i$-th I/O. The requirements on the present FPGA are that many strips, or pixels, belonging to a position sensitive detector (PSD) are analysed using a single FPGA. Moreover, for an optimal discrimination of the real neutron signals from the whole of the other signals, due to either electronic or ambient noise or to $\gamma$ radiation, some analysis of the signal is necessary.

A first analysis can be based on a filtering of the signal to get a properly shaped pulse. The amplitude of the pulse is proportional to the energy released after the neutron capture inside the detector, as it is done using usual hardware front-end electronics. Filtering, usually accomplished by the detector preamplifier, can be obtained through a digital elaboration of the sampled signal. For example, a simple low-pass filter is formally obtained in the time domain by the following integral:

\begin{equation}
v_u(t) = {1 \over \tau} \, \int_{-\infty}^{t} \, v_i(t) \, \exp\Big[{t' - t \over \tau}\Big] \, d t'
\end{equation}

\noindent
where $v_i(t)$ and $v_u(t)$ are the input and output signals respectively and $\tau$ is the decay time of the filter.

A general linear filter is simply given by:

\begin{equation}
v_u(t) = \int_{-\infty}^{t} \, v_i(t) \, T_f(t - t') d t'
\end{equation}

\noindent
where $T_f(t)$ is the transfer function of the filter. If the signals are sampled with a period $T$, the integral can be transformed into a series that, in practical cases, reduces to a finite sum:

\begin{equation}
v_u(t_k) = \sum_{i=-\infty}^\infty \, T_f^d(t_k,t_i) \, v_i(t_i)
\end{equation}

\noindent
where $t_k$ is the running sampling time and $T_f^d(t_k,t_i)$ is the transfer matrix. In the case of a low-pass filter, the transfer matrix is readily obtained:

\begin{equation}
T_f^d(t_k,t_i) = \cases{\exp\Big[{t_i - t_k \over \tau} \Big] [1 - \exp\Big({-T \over \tau} \Big)],&if $t_i \le t_k$ \cr
	0,&otherwise \cr}
\end{equation}

\noindent
where $T$ is the sampling time. 
In real cases the series must be substituted by a finite sum in a way that $t_k - t_i \gg \tau$. This equation is useful for practical purposes. Interestingly, low order filters are relatively simple to be implemented as to hardware and numerical solutions. On the contrary, the hardware implementation of high order filters is practically impossible, although numerical solutions are as simple as low order filters.
We also note that all the analog linear electronic filters are infinite impulse response (IIR), namely the output signal lasts ideally forever after the arrival of a short pulse at the input. On the contrary, almost all the digital filters are finite impulse response (FIR), that is the output signal lasts for a given period after the arrival of a short pulse at the input. Therefore, a real time digital analysis can use FIRs practically of any shape.

For real applications, it can be important to introduce more complex data treatments to extract the cleanest information from the signal produced by the detector. A better data processing is certainly  more relevant in the case of solid state detectors where the charge collection at the detector is fairly low and the electronic noise becomes an important issue, in addition to the point of $\gamma$-ray rejection. Also a specifically tailored data treatment can be valuable in the case of scintillator detectors coupled to SiPMs for scintillator light collection. SiPMs~\cite{sipm} are actually very promising devices with a potential to replace vacuum phototubes because of their very small size, good performances under vacuum and high magnetic fields, the negligible heat release and the low cost for large production. Their limitation is, however, the very high thermal noise that can mask the neutron signal, particularly when low light emission lithium glass scintillators are employed.

\begin{figure}
%\vspace{20mm}
\includegraphics[width=0.9\textwidth,keepaspectratio]{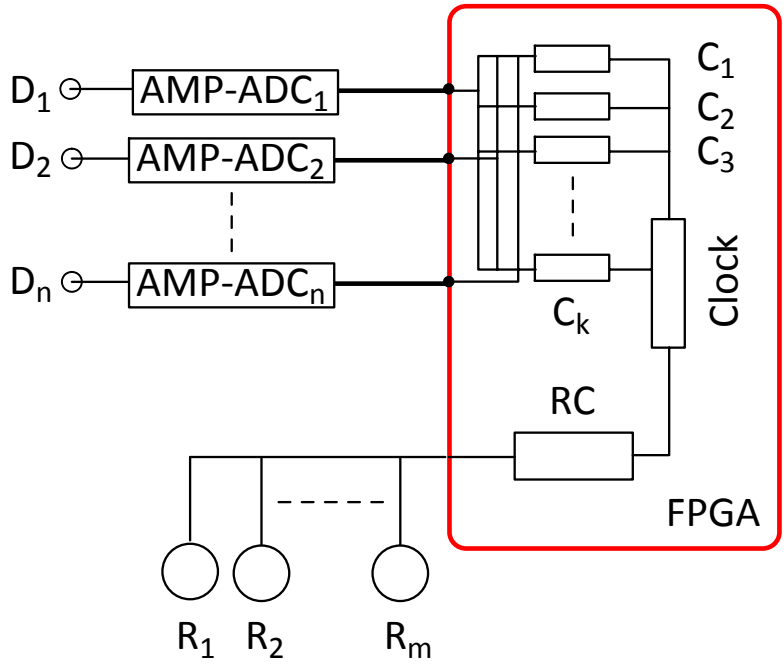}
\caption{Block description of the proposed acquisition system for real time pulse shape analysis and single event acquisition, other than control of ToF components. The $n$ blocks AMP-ADC are wide band amplifiers followed by the analog to digital converters. The front end electronics is not shown because it is specifically related to the detector under use. The FPGA is programmed in such a way to have specific calculation elements $C_1 ... C_k$ to perform the different steps of the pulse shape analysis. The RC is the rotation control system used to drive different components. The Clock is the same for elements.}
\label{fig6}
\end{figure}

To increase the flexibility of the system we introduced a procedure to perform all the data treatment by special programming the FPGA. The block diagram of the proposed system is shown in Fig.~\ref{fig6}. A wide band, low noise linear amplifier is used to amplify the incoming voltage signal. The so amplified output signal is used to feed a fast enough ADC so that the parallel 8 bits output can be sent to the FPGA that is used also to provide the clock of the ADC. A full analysis of the sampled detector signal is then performed by the FPGA to obtain a real time detection of the true neutron signals together with their arrival time data. Several calculation units are used to perform, in real time, a pulse shape analysis. The same FPGA can also be employed to control other units of the instrument relevant for the ToF acquisition, as it was demonstrated by the approach applied to control and collect the data in the rotating crystal monochromator test~\cite{rotat}. Considering the constraints of a real time pulse shape analysis on single neutron events, we are additionally developing a new architecture implemented on an FPGA to perform, on a single device, several activities necessary to produce the correct information for a set of neutron events. This system, which is based on the implementation of several specialised digital processors on the FPGA to perform the appropriate analysis, will be presented in detail in a subsequent pubblication~\cite{bmac}.

The effectiveness of the proposed pulse shape analysis was tested first by computer simulation and then making neutron tests with two different devices, namely a silicon PIN diode 0.3 mm thick coupled to a Gd$_2$O$_3$ converter, and a SiPM with a GS20 scintillator lithium glass scintillator. The PIN was coupled to a front-end electronics that enabled the use the output signal also in the standard way by a window discriminator. The SiPM had no front-end electronics as the signal was provided by the voltage on a 50 $\Omega$ resistor installed at the end of a coaxial cable. The front-end electronics of the PIN diode (Hamamatsu, S3590-09, 1 x 1 cm$^2$ area) consisted of a low noise A225 preamplifier (Amptek) followed by a linear amplification stage (voltage gain = 100). The Gd$_2$O$_3$ neutron converter was a 3-10 $\mu$m thick coating on a 2 $\mu$m thick mylar foil to make the converter removable. The output was digitised using a pico scope 2000 (Pico Technology) with a bandwidth of 100 MHz and selecting a sampling time $T = $100 ns. The output of the SiPM (FBK Trento, 3 x 3 mm$^2$) was digitised using the same device but with a sampling time $T = $1 ns. Application of this digital acquisition system is convenient for the testing phase, because the sampled neutron signal can be stored for the subsequent analysis that simulates different form of real time acquisition using the same data.

The present approach relies on a full pulse shape analysis, instead of turning to a filter to identify the correct signals. We defined the correct shape of the reference neutron signal $V_r(t)$ for comparison with the measured signal $V_m(t)$, sampled on $N$ time points $t_i$. The comparison can be carried out in real time if the sequence $V_m(t_i), t_i, i = 1,...,N$ is stored into an appropriate buffer. This is easily carried out in the simulation phase but, for real time applications, a proper procedure has to be implemented on an FPGA. The following operation can be used:

\begin{equation}
A_s = {\sum_i \, V_m(t_i) \over \sum_i \, V_r(t_i)}
\end{equation}

Obtaining $A_s$ implies just two simple operations, once the fixed reference signal $V_r(t)$ is stored in a given buffer. Indeed, one has to calculate $\sum_i \, V_m(t_i)$ and then the ratio to the fixed sum $\sum_i \, V_r(t_i)$. A real time application can be simplified by a proper choice of $V_r(t)$ such that $\sum_i \, V_r(t_i) = 2^n$, so that the calculation of the ratio $\sum_i \, V_m(t_i)$ over $\sum_i \, V_r(t_i)$ is much faster. To establish if the sequence $V_m(t_i), t_i, i = 1,...,N$ refers to a genuine neutron signal, one can search for sequences such that:

\begin{equation}
S = \sum_i \, [ V_m(t_i) - A_s \, V_r(t_i)]^2 < S_o
\end{equation}

Holding the above relationship, it can be assumed that the sequence starting at $t_1$ indicates the collection of a neutron in the detector at the time $t_c$, which can be deduced from the reference signal $V_r(t)$. The condition is quite strong and, once $V_r(t)$ and $\chi^2_0$ are properly chosen, the procedure is much more efficient than all possible hardware circuits, having full flexibility when implemented on a FPGA.

\begin{figure}
%\vspace{20mm}
\includegraphics[width=0.9\textwidth,keepaspectratio]{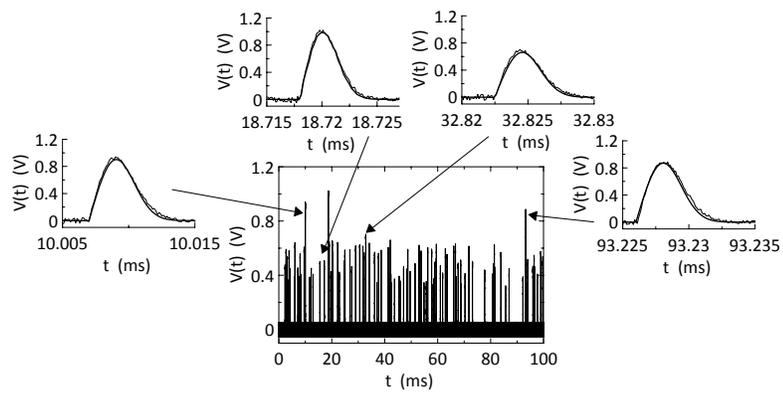}
\caption{A typical signal produced in the simulation, lasting 100 ms, is presented. Almost all the visible peaks do not belong to a nominal neutron peak. The four neutron peaks present in the signal are shown in the expanded plots (thin line) in comparison with the model fitting curves (thick line).}
\label{fig7}
\end{figure}

As described, the outlined procedure was tested using simulated signals, randomly generated as to arrival time and amplitude and slightly fluctuating pulse length. A Gaussian random noise was added to the so generated signals to simulate a real output from the present detector system. Additional peaks with a slightly different shape were added to the data to simulate signals produced by possible background radiation. The simulation enabled to identify the error level of the analysis based on the described approach. Indeed, the correct neutron signals were located at the known time positions and the wrong signals were identified as well. The results of the simulation are shown in Fig.~\ref{fig7} where a typical sampled signal is shown together with the neutron peaks fitted by the equation below:

\begin{equation}
V_r(t) = {1 \over 2} \Big \{1 + \tanh\Big[{t - t_o \over \tau}\Big] \Big\} \cdot \exp[-b (t - t_o)^2]
\end{equation}

\noindent
where $t_o$ is the arrival time, $b = 170000$ ms$^{-2}$ and $\tau = 0.001$ ms for the peaks that can be attributed to neutrons. The simulated sequences lasted 0.1 s and the sampling frequency was 10 MHz. A total number of 4 neutrons is detected in the typical data shown in Fig.~\ref{fig7}. These 4 neutron peaks correspond to a 100\% efficiency of the analysis. No wrong signal was picked up by the analysis procedure even though the number of spurious peaks was 104, that is much higher than the number of neutrons. To complete the simulation, several different sequences were produced in order to get better statistics. The error level estimated with this approach is $\approx 7 \cdot 10^{-5}$ with only 5 wrong signals over more than 72400 peaks analysed. We note that the present selection efficiency is not the detection efficiency of the device, as here the output signal from the amplifier is analysed. In conclusion, according to the results of the simulations, we are quite confident that the outlined procedure can help in obtaining a good real time discrimination of the neutrons even in the worse environment and working conditions of the detector.

\begin{figure}
%\vspace{20mm}
\includegraphics[width=0.9\textwidth,keepaspectratio]{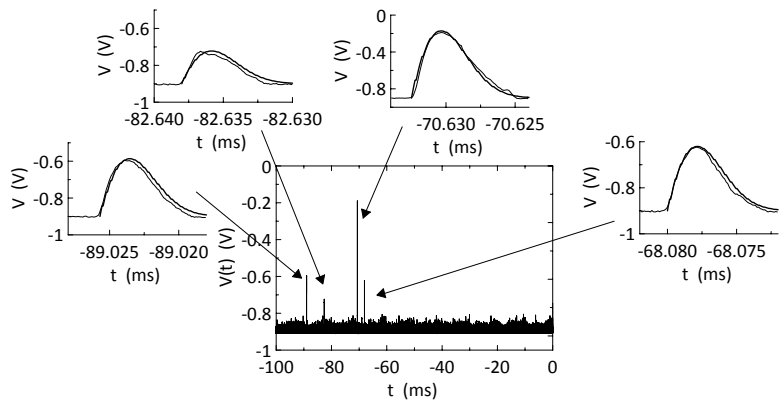}
\caption{Typical experimental spectrum collected from a PIN diode plus Gd$_2$O$_3$ converter. The signal contains four peaks identified as neutrons clearly visible above random noise. The four neutron peaks present in the signal are shown in the expanded plots (thin line) in comparison with the model fitting functions (thick line).}
\label{fig8}
\end{figure}

The PIN and SiPM experimental results were analysed using the same procedure. The data shown in Fig.~\ref{fig8} refers to the measurements on the Si PIN diode coupled to $^{157}$Gd$_2$O$_3$ converter layer. The data were collected on the three-axis spectrometer IN3 at the ILL. The neutron capture inside the Gd layer is followed by $\gamma$-ray emission due to the de-excitation of the nuclei. For each absorbed neutron one electron with an energy ranging from 20-30 keV up to about 250 keV is emitted with a probability of $\approx$ 80 \%. According to this scheme, and based on the analysis approach, we evaluate an efficiency of $\sim$30\% for the present PIN detector. The result obtained is quite promising, especially considering that the deposition technique of this converter is still not optimized and improvements are possible. Indeed, the uniformity of the converter layer is a primary feature for the detector to work properly and an efficiency decrease can be due to a non-uniform deposition layer. 

The case of the SiPM is much more demanding in terms of signal analysis because of the very high noise and the rather low light output of the GS20 scintillator. Indeed, the detector should be self-triggering and simple filtering techniques are hardly effective in reducing the noise. Therefore, pulse shape analysis was preliminarly carried out to understand the effectiveness of such an approach in the real time application. The GS20 scintillator, 1.5 mm thick, had the same lateral size as the SiPM and was coupled to it by a 3 mm thick plexiglass block. A teflon tape, all around the system, worked as a light reflector. The detector was tested on the monochromatic beam of the IN3 three axis spectrometer at ILL with a selected wavelength of 2.4 \AA. 
Placing the detector on the primary beam provided enough intensity for all the test configurations, although background could be higher. The detector was shielded by a 5 mm thick elastobore, which provides a nominal attenuation by a factor 10$^3$ at thermal wavelength. The expected intensity impinging the detectors was about 10$^7$ n/cm$^2$ s, so that about 10$^6$ n/ s were expected on the detector area. As this rate is rather high, the incoming beam was attenuated by means of a plexiglas plate, 8 mm thick, which gives an intensity of the order of 2$\cdot$10$^4$ n/s on the SiPM detector. This intensity is appropriate for a test of this detector. Indeed, it is expected to be rather fast because its dead time is mainly limited by the primary decay time of the GS20 scintillator. 

As the SiPM is a rather new device for use as a neutron detector~\cite{sipm}, and because of the relatively high environmental background and the intrinsic background sources, the shape analysis was carefully performed. Different tests were carried out by changing the SiPM overvoltage and the position of the detector in the beam to take into account the not uniform intensity distribution of the beam, a problem that cannot be ignored when testing a small detector like this one.

\begin{figure}
%\vspace{20mm}
\includegraphics[width=0.9\textwidth,keepaspectratio]{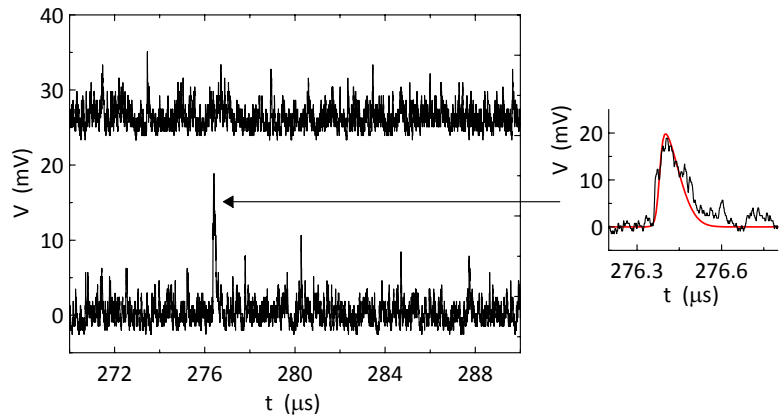}
\caption{Typical experimental spectrum collected from the SiPM coupled to a GS20 scintillator (bottom black line). The signal, lasting 1 ms, is presented on a fraction of the total duration to enhance the presence of a peak we can attribute to neutrons signal by comparison to the beam-off spectrum (top black line). The enlarged plot on the right side shows the peak we attribute to neutrons in comparison with the fitted peak (red thick line).}
\label{fig9}
\end{figure}

\begin{figure}
%\vspace{20mm}
\includegraphics[width=0.9\textwidth,keepaspectratio]{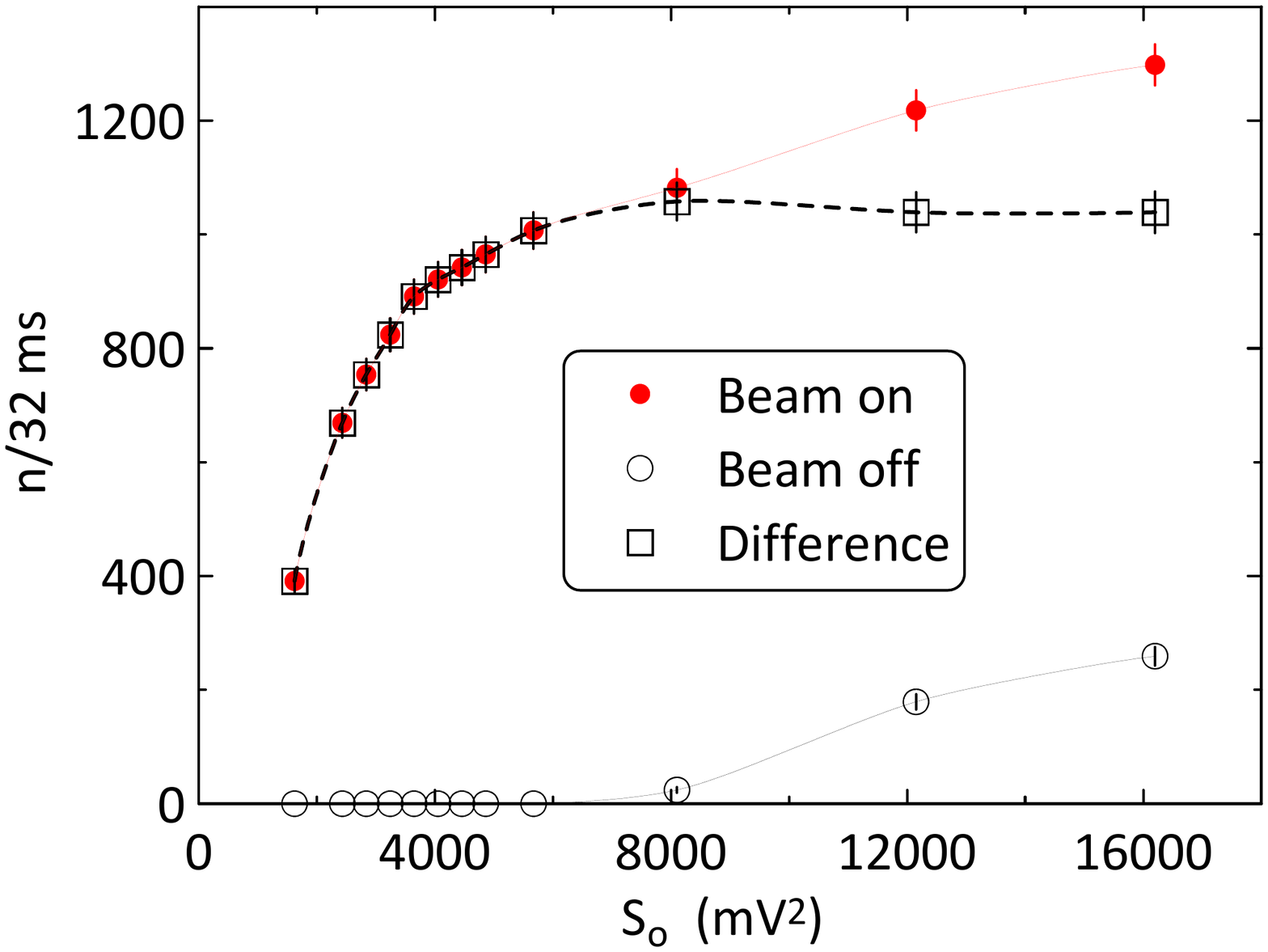}
\caption{Results of the full pulse shape analysis. Neutron flux as a function of $S_o$. Red full circles, beam-on data. Black open circles, beam-off data. Black empty squares, difference between beam-on and beam-off data. The lines are guide to eyes.}
\label{fig10}
\end{figure}

A typical signal at the SiPM load resistor is shown in Fig.~\ref{fig9}. The effect of the noise can be noticed as well as the appearance of some peaks. A neutron event can be attributed to peaks exceeding 12-13 mV, as, indeed, they are rare when the neutron beam is switched off. From several similar acquisitions, all lasting 100 ms, we selected some of the highest peaks. It was then possible to obtain reasonably good fits of the peaks attributed to neutrons, using the same simple function employed in the case of the PIN diode. Of course, the parameters of the fitting function are very different as the SiPM is a fast device with no front-end electronics and the data were collected using a short sampling time (1 ns). The analysis of the data was found to work well. Several tests were performed, in particular, the data were analysed by using a low- and a high-level threshold and by varying $S_o$. After various trials, we set the low level threshold at 9 mV and the high level threshold at 40 mV. Finally, a set of 32 acquisitions, lasting 1 ms each, were stored with beam open and beam closed. Both sets of data were analysed applying the same procedure, to isolate the contribution coming from noise and environmental background. The results are shown in Fig.~\ref{fig10}, where we plot the number of peaks identified as neutrons as a function of $S_o$. It is expected that too a low $S_o$, while rejecting all the background, produces also a rejection of neutron counts. By subtracting the beam-off background from the beam-on data, it was easy to identify the best choice of $S_o$ to get almost no background counts and about 3$\cdot$10$^4$ n/s with the beam on. This intensity value is consistent with the expected intensity, but we had no means to check it on an absolute scale. We can only observe that the scintillator has a nominal efficiency close to 100 \% and the experimental results suggest a light selection efficiency also close to 100 \%. 

To state that a SiPM based system is adequate as a neutron detector on a real neutron instrument, more tests need to be done. Also, every potential use of this prototype on a larger scale requires new effort in designing and testing proper front end electronics.

\section{Conclusions}

Solid-state detector technology is nowadays a possible alternative in neutron detection which can find application in future instrumentations. The promising performances and the capability to couple high integration electronics push forward the development of real time and single event analysis systems. This is a significant advance bringing in the potential to achieve higher performances and to meet the requirements of future applications in high brilliance neutron beams. Here we presented the results obtained by testing three different Si-based devices and discussed a pulse shape analysis approach that was applied to both simulations and measurements, with the aim of developing a real time data analysis system.
\\ Si microstrip based position sensitive neutron detectors can achieve sub-millimiter spatial resolution and high counting rate, with an efficiency currently of $\approx 25$ \% down to about 1.5 \AA. This technology can be a suitable alternative for applications that require high spatial resolution and counting rate capability, i.e. neutron reflectometry. 
\\ Si PIN diodes with an efficiency of $\approx 30$ \% have been tested. A promising increase is expected by the optimisation of the deposition technique for the $^{157}$Gd$_2$O$_3$.
\\ The SiPM device described in this paper can reach a spatial resolution of $\approx 1$mm$^2$ and counting rate above 500 kHz (with 10 \% of dead time).  
\\The pulse shape analysis method here proposed fits well for these detector technologies and its application leads to improved performances.
Furthermore, the approach can be implemented on FPGA, adopting simplified numeric calculation as well. A detail description of the method will be discussed in a following publication~\cite{bmac}.

\newpage

\noindent\rule{50mm}{0.2mm} \\

ACKNOWLEDGEMENTS

The ILL and ISIS neutron sources are kindly acknowledged for providing the neutron beam time. Martin Boehm and Andrea Piovano (ILL), and Nigel Rhodes (ISIS) are acknowledged for the support during the neutron tests. This research project has been partially supported by the European Commission under the 7th Framework Programme through the activity {\it Integrated Infrastructure Initiative for Neutron Scattering and Muon Spectroscopy (NMI3)}, contract No. 226507.

\end{document}